\documentclass[%
 reprint,
 amsmath,amssymb,
 aps,
 prb,
]{revtex4-1}

\usepackage{xcolor}
\usepackage{graphicx}
\usepackage{dcolumn}
\usepackage{bm}
\usepackage{hyperref}

\begin{document}

\preprint{APS/123-QED}

\title{Quantum phase transitions in the alternating XY chain with three-site interactions}

\author{Kaiyuan Cao}
\affiliation{Zhejiang Lab, Hangzhou 311100, P. R. China}

\author{Hao Fu}%
\affiliation{Department of Physics and Institute of Theoretical Physics, Nanjing Normal University, Nanjing 210023, P. R. China}

\author{Xue Liu}%
\affiliation{Department of Physics and Institute of Theoretical Physics, Nanjing Normal University, Nanjing 210023, P. R. China}

\author{Ming Zhong}%
 \email{mzhong@njnu.edu.cn}
\affiliation{Department of Physics and Institute of Theoretical Physics, Nanjing Normal University, Nanjing 210023, P. R. China}

\author{Peiqing Tong}
 \homepage{pqtong@njnu.edu.cn}
\affiliation{Department of Physics and Institute of Theoretical Physics, Nanjing Normal University, Nanjing 210023, P. R. China}
\affiliation{Jiangsu Key Laboratory for Numerical Simulation of Large Scale Complex Systems, Nanjing Normal University, Nanjing 210023, P. R. China}

\date{\today}

\begin{abstract}
  We investigate the quantum phase transition in the alternating XY chain with the XZX+YZY type of three-spin interactions. We present the exact solution derived by means of the Jordan-Wigner transformation and study the average magnetization, spin correlations, and von Neumann entropy to establish the phase diagram. The phase diagram consists of the ferromagnetic phases, the paramagnetic phases, and the phase with weak magnetization (WM). By examining the nearest-neighbor transverse spin correlation, we probe that in the WM phase, the spins within a supercell generate a cluster with a small total spin, but between the nearest-neighbor supercells are distributed randomly. Especially for the dimerized limit case, the spins within a supercell tend to point to opposite directions of the transverse field. In addition, we also investigate the influence of the three-site interaction, and find that the WM phase is absent as the strength of the three-site interaction increases. Our findings shed light on the complex behavior of the alternating XY chain and provide valuable insights for future studies.
\end{abstract}

\maketitle


\section{Introduction}

The system of interacting spins on a lattice provides an incredibly versatile platform for investigating intricate quantum phases and phase transitions that arise due to the interplay of various quantum phenomena and symmetries \cite{sachdev2011, Diep2013}. Various forms of spin interaction are realized in lattice spin systems, which have discovered a plethora of quantum phases, such as symmetry-protected topological phases \cite{Senthil20156, Ryu2015, Song2015, Wen201789} and quantum spin liquids \cite{Anderson1973, Wen201789, Wen200265, Patrick200678, Patrick2008321, Balents2010464, Savary201780}. In addition to the spin interaction, with the rapid development of spin-orbit physics, the strong spin-orbit coupling in the magnetic systems leads to intrinsically frustrated orbital (pseudospin) interactions. Such pseudospin interactions are different from the Heisenberg SU(2) isotropic exchanges and have attracted much attention from both theoretical \cite{Zhao2008100, Wu2008100, Chaloupka2010105, Wohlfeld2011107, Balents20145, Brzezicki20155, Hermanns20189} and experimental \cite{Kim2009323, Galitski2013494, Takagi20191, Broholm2020367} perspectives. In this regard, studying interacting lattice spin systems can not only help us understand the complexity of quantum phases and phase transitions but also provide an important theoretical basis for exploring new quantum materials and applications.

Most lattice spin models with complex interaction are difficult to solve rigorously. A sensible and common approach is to extend a model known to be exactly solvable by adding additional interactions or modifying its lattice. This method has repeatedly achieved great results. For instance, the chiral phase is found in the XY chain with the additional Dzyaloshinskii-Moriya interaction \cite{Siskens197579, Perk197658, Derzhko200673, Jafari200878}, which can be used to explain the anti-ferromagnetic behavior \cite{Dzyaloshinsky19584, Moriya1960120}; The detailed localized-extended classification phase diagram of ferromagnetic (FM) and paramagnetic (PM) phases was explored in the transverse field Ising model in a quasi-periodic lattice \cite{Chandran20177, Crowley2018120, Divakaran201898}. The later example of the quasi-periodic model has also been investigated using the asymptotic results of the periodic chain in the increasing-size superlattice\cite{Tong2001304, Tong200265, Tong200697}. It is worth noting that a unique magnetic phase, characterized by zero magnetization, arises under certain periodic modulations. However, this phase has not received significant attention in previous studies, leading to a limited understanding of its properties. Despite being classified as the paramagnetic phase \cite{Zhong201043}, there is a lack of research investigating the local order parameter of this phase.

Additionally, the literature focuses on the nearest-neighbor interactions in most systems. However, considering the spin interaction originating from the superexchange of the long-range Coulomb potential, the effect of long-range interactions must be addressed when applying the theory to realistic materials. Recently, three-site interactions received considerable attention\cite{Gottlieb199960, Titvinidze200332, Lou200470, krokhmalskii200877, Cheng201081, Derzhko201183, Li201183, Song2015, Liu201201, Menchyshyn201592, You201693}. For instance, the chiral phase is also found in the quantum spin chains with the XZY-YZX type of three-site interaction \cite{Liu201201, You201693}. Two kinds of spin liquid phases are present in the anisotropic XY chain with XZX+YZY type of three-site interaction \cite{Titvinidze200332, krokhmalskii200877, Menchyshyn201592}. Such complex interactions between three subsequent sites essentially enrich the ground state phase diagram of the spin model and open new opportunities for underlying physics.

\begin{figure}
    \centering
    \includegraphics[width=1\linewidth]{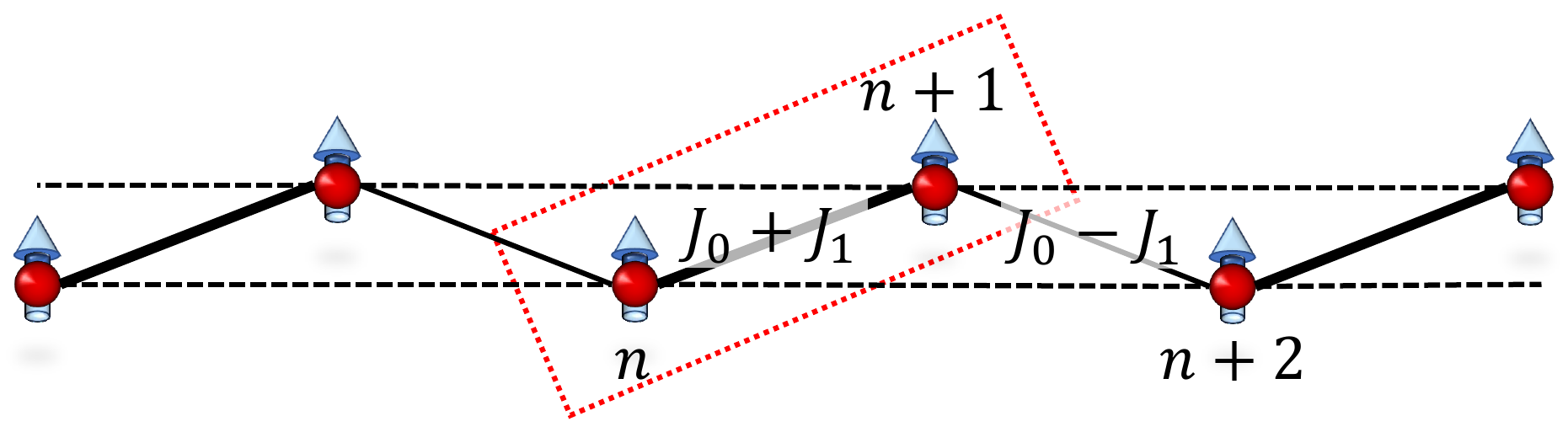}
    \caption{Schematic representation of the structure of the alternating XY chain.}
    \label{fig:schematic-structure}
\end{figure}

In this paper, we consider the alternating XY chain \cite{Perk1975319, Siskens197521} with XZX+YZY type of three-site interaction in the transverse field [see Fig.~\ref{fig:schematic-structure}], which is exactly solvable. A similar model was studied by Zvyagin in Ref.~\onlinecite{Zvyagin200980} before, but the phase diagram of the current model is still unclear and requires further investigation to understand the properties of each phase. By calculating the average magnetization, spin correlations, and von Neumann entropy, we obtain the ground state phase diagram, consisting of the ferromagnetic phases, the paramagnetic phases, and the phase with weak magnetization (WM). By examining the nearest-neighbor transverse spin correlation, we find that in the WM phase, spins within a supercell generate a cluster with a small total spin, but between the supercells are distributed randomly. The total spin of a cluster in a supercell is dependent on the dimerized structure, denoted by the ratio $\frac{J_{0}-J_{1}}{J_{0}+J_{1}}$. In particular, we observe that for the dimerized limit case, the spins within a supercell tend to point to opposite directions of the external field. This is the reason for the weak magnetization (or even zero magnetization in the fully dimerized case) in the WM phase. Moreover, by investigating the von Neumann entropy, we notice that the von Neumann entropy in the WM phase has certainly large value which is different from the cases in the paramagnetic phase. The WM phase induced by the alternating interaction thus should not be classified as the PM phase. In addition, we also discuss the influence of the three-site interaction, and find that the WM phase is absent as the strength of the three-site interaction increases, instead of the appearance of two ferromagnetic phases.

The article is organized as follows. In Sec.~II, we introduce the alternating XY chain with XZX+YZY type of three-site interaction. By exactly diagonalizing the Hamiltonian, we obtain the expression of the quasiparticle excitation spectra and the phase diagrams separated by the quantum critical lines. In Sec.~III, the average magnetization and spin correlations are calculated to identify each phase. In Sec.~IV, we study the behavior of von Neumann entropy within each phase and on the quantum critical lines. Finally, we give a summary of our main results and conclude in Sec.~V.

\section{Model}

The Hamiltonian of the one-dimensional extended XY chain in the transverse field with the alternating hopping and XZX+YZY type of three-site interaction is
\begin{equation}\label{eq: Hamil}
    \begin{split}
        H & = -\frac{1}{2}\sum_{n=1}^{N}\{\frac{J_{0}+(-1)^{n-1}J_{1}}{2}[(1+\gamma)\sigma_{n}^{x}\sigma_{n+1}^{x} \\
          &  \quad\quad\quad +(1-\gamma)\sigma_{n}^{y}\sigma_{n+1}^{y}] + h\sigma_{n}^{z} \} \\
          &  \quad -\frac{1}{2}\sum_{n=1}^{N}\Omega (\sigma_{n-1}^{x}\sigma_{n}^{z}\sigma_{n+1}^{x}+\sigma_{n-1}^{y}\sigma_{n}^{z}\sigma_{n+1}^{y}),
    \end{split}
\end{equation}
where $\sigma^{x,y,z}$ are the Pauli matrices, $J_{n}=J_{0}+(-1)^{n-1}J_{1}$ are the strength of the hopping interactions between the nearest-neighbor spins, $\gamma$ is the anisotropic parameter, and $h$ is the external field. The Hamiltonian (\ref{eq: Hamil}) can be transformed to the extended XY chain in the alternating field $[h_{n}=h_{0}+(-1)^{n-1}h_{1}]$ due to the duality transformation \cite{Suzuki2013, Chandran20177}. For convenience, we define the parameter
\begin{equation}
    \alpha = \frac{J_0-J_1}{J_0+J_1}
\end{equation}
with $J_0+J_1=1$ without losing the generality. Assuming periodic boundary conditions, the Hamiltonian (\ref{eq: Hamil}) can be exactly solved using the Jordan-Wigner transformation that maps the Hamiltonian (\ref{eq: Hamil}) to a system of spinless fermion in the double lattice. Since each momentum is decoupled, we obtain the diagonalized Hamiltonian
\begin{equation}
    H = \sum_{k} [\Lambda_{k1}(\eta_{k1}^{\dag}\eta_{k1}-\frac{1}{2}) + \Lambda_{k2}(\eta_{k2}^{\dag}\eta_{k2}-\frac{1}{2})]
\end{equation}
in momentum space after performing the Fourier and Bogoliubov transformations, where the Bogoliubov coefficients are obtained by solving the eigenproblem of the Bloch Hamiltonian $\mathbb{H}_{k}$ (more details seen in Appendix~A). The eigenproblem of the Block Hamiltonian $\mathbb{H}_{k}$ is equivalent to solving a quartic equation concerning the quasiparticle excitation spectra $\Lambda_{k1(2)}$. The solutions are given by
\begin{eqnarray}
    \Lambda_{k1} &=& \sqrt{\frac{P-\sqrt{P^{2}-4Q}}{2}}, \\
    \Lambda_{k2} &=& \sqrt{\frac{P+\sqrt{P^{2}-4Q}}{2}}
\end{eqnarray}
with
\begin{equation}
    P = 2|A_{k}|^{2} + 2|B_{k}|^{2} + 2(h-\Omega\cos{k})^{2},
\end{equation}
\begin{equation}
    \begin{split}
        Q & = (h-\Omega\cos{k})^{4} + 2(h-\Omega\cos{k})^{2}(|B_{k}|^{2}-|A_{k}|^{2}) \\
          & \quad + (B_{k}^{2}-A_{k}^{2})[(B_{k}^{*})^{2}-A_{k}^{*})^{2}],
    \end{split}
\end{equation}
\begin{equation}
    A_{k} = \frac{1}{2}(1+\alpha e^{ik}), \quad \text{and} \quad B_{k} = \frac{1}{2}\gamma(1-\alpha e^{ik}).
\end{equation}

\begin{figure*}
    \centering
    \includegraphics[width=1\linewidth]{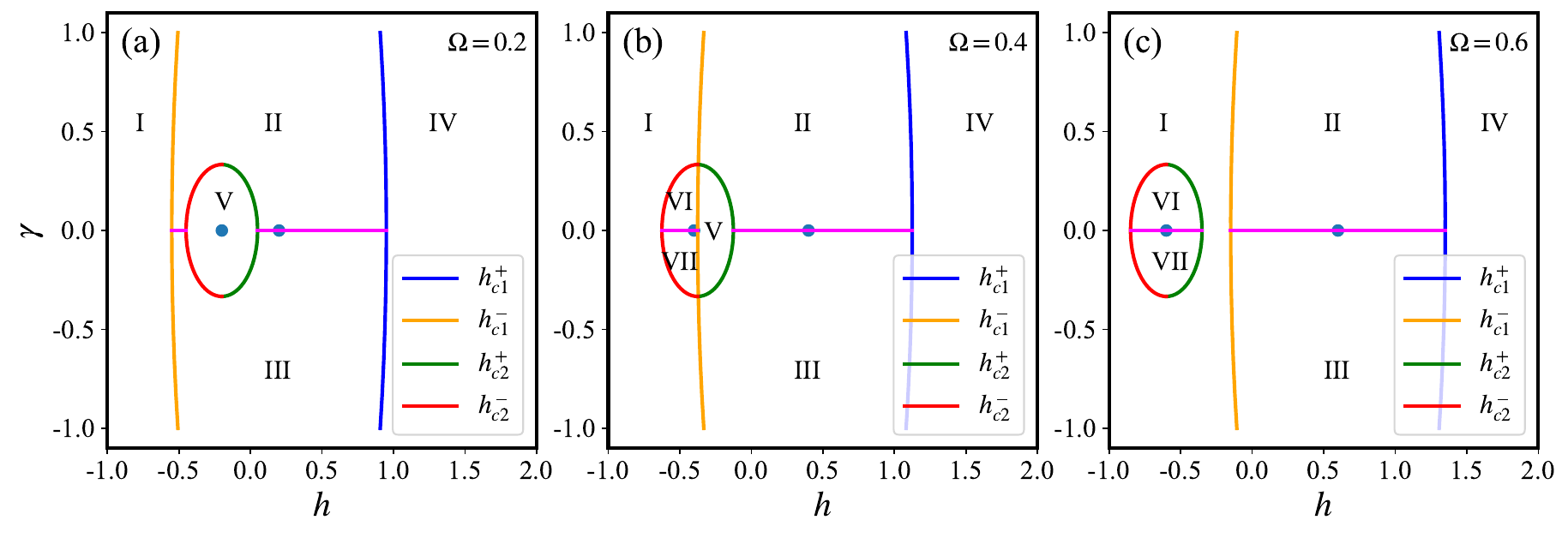}
    \caption{Phase diagrams in the $(h-\gamma)$ plane with fixed $\alpha=0.5$ for (a) $\Omega=0.2$, (b) $\Omega=0.4$, and (c) $\Omega=0.6$, respectively.   }
    \label{fig:phase-diagram}
\end{figure*}

\begin{figure}
    \centering
    \includegraphics[width=1\linewidth]{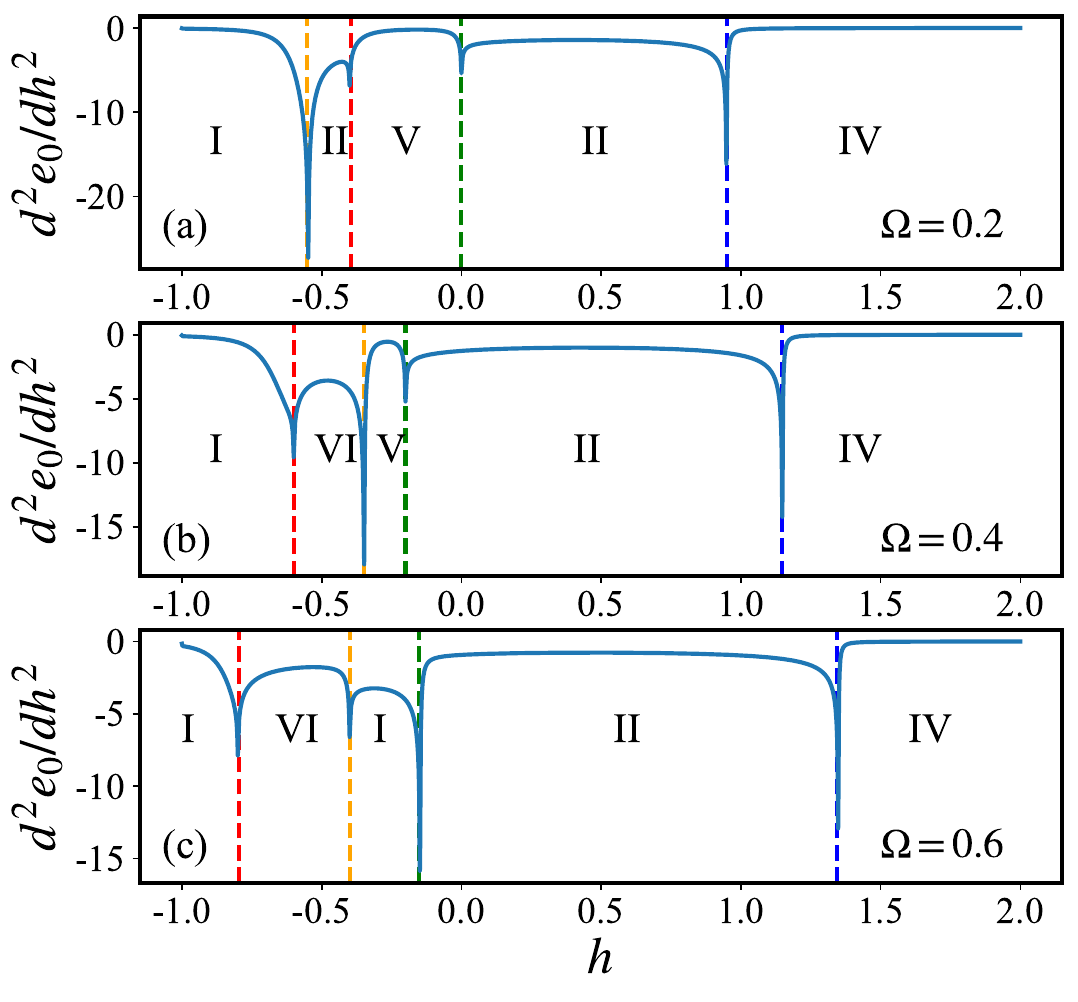}
    \caption{The second-order derivative of the average ground-state energy $e_{0}$ as a function of the external field $h$ for (a) $\Omega=0.2$, (b) $\Omega=0.4$, and (c) $\Omega=0.6$, with fixed $\alpha=0.5$ and $\gamma=0.2$. }
    \label{fig:average-energy-h}
\end{figure}

\begin{figure}
    \centering
    \includegraphics[width=1\linewidth]{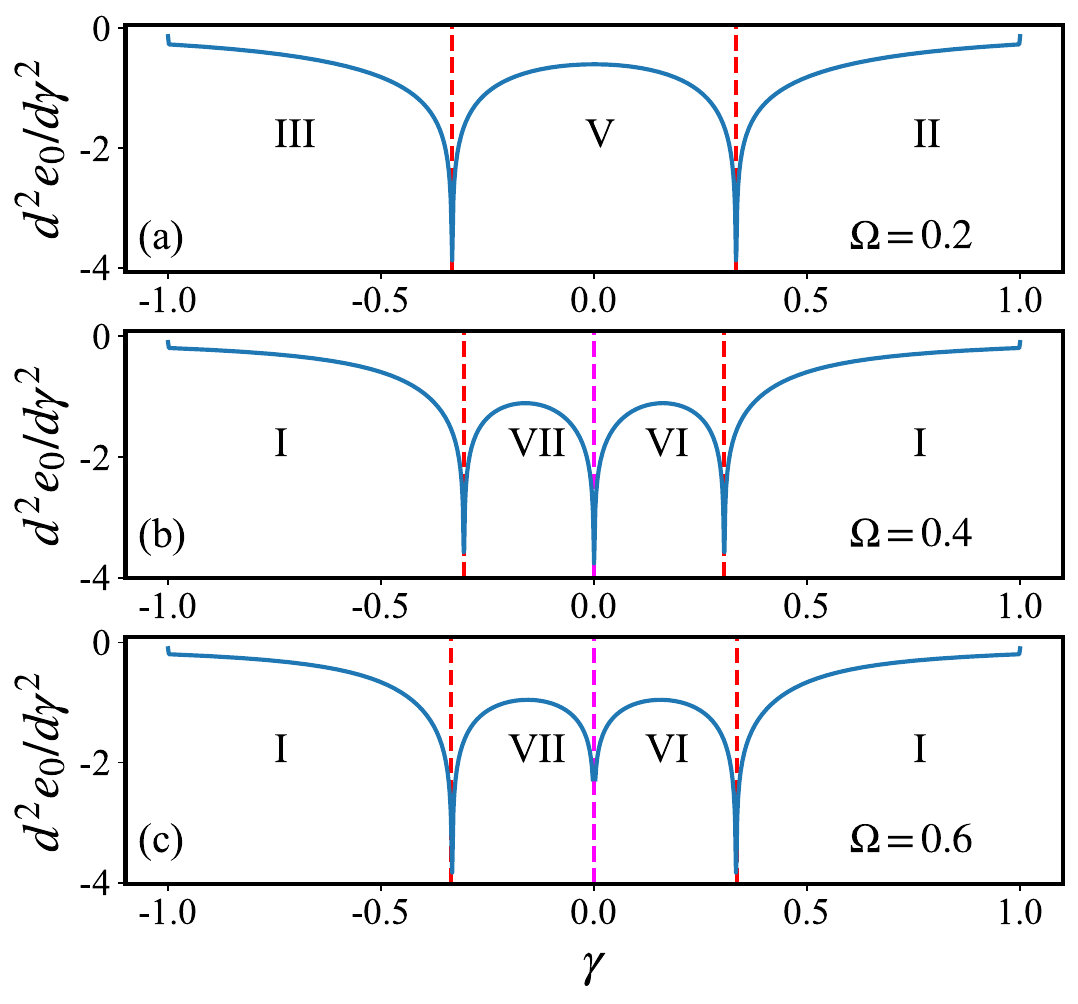}
    \caption{The second-order derivative of $e_{0}$ as a function of the anisotropic parameter $\gamma$ for (a) $\Omega=0.2, h=-0.2$, (b) $\Omega=0.4, h=-0.5$, and (c) $\Omega=0.6, h=-0.6$.}
    \label{fig:average-energy-gamma}
\end{figure}

The critical point of the QPT appears where the band gap closes, corresponding to $\mathrm{min}(\Lambda_{k1})=0$. Hence, we obtain the critical point by solving the minimum value of the lower quasiparticle excitation spectrum, i.e.
\begin{equation}
    \left\{\begin{array}{r}
             \Lambda_{k1} = 0, \\
             \frac{d\Lambda_{k1}}{dk} = 0.
          \end{array}
    \right.
\end{equation}
With fixed $\alpha$ and $\Omega$, we obtain the critical lines in the $(h-\gamma)$ plane: for $-1\leq\gamma\leq1$,
\begin{equation}\label{eq:critical-line-1}
    h_{c1}^{\pm} = \Omega \pm \frac{\sqrt{(1+\alpha^{2})(1-\gamma^{2})+2\alpha(1+\gamma^{2})}}{2};
\end{equation}
for $-\frac{1-\alpha}{1+\alpha}\leq\gamma\leq\frac{1-\alpha}{1+\alpha}$,
\begin{equation}\label{eq:critical-line-2}
    h_{c2}^{\pm} = -\Omega \pm \frac{\sqrt{(1+\alpha^{2})(1-\gamma^{2})-2\alpha(1+\gamma^{2})}}{2}.
\end{equation}
According to Eqs.~(\ref{eq:critical-line-1}) and (\ref{eq:critical-line-2}), we know that the distance of critical lines $h_{c1}^{\pm}$ for $\gamma=0$ is $1+\alpha$, the center of the critical lines $h_{c1}^{\pm}$ is $\Omega$, the distance of $h_{c2}^{\pm}$ for $\gamma=0$ is $1-\alpha$, and the center of critical lines $h_{c2}^{\pm}$ is $-\Omega$. We thus obtain the phase diagram, and it is foreseeable that the model exhibits a non-symmetric phase diagram with respect to $h=0$ due to the three-spin interactions, as three typical instances shown in Fig.~\ref{fig:phase-diagram}:
\begin{itemize}
    \item If $\Omega<\frac{\alpha}{2}$, the critical line $h_{c2}^{-}$ is located at the left of $h_{c1}^{-}$, and the phase diagram consists of five parts [see regions I, II, III, IV, and V in Fig.~\ref{fig:phase-diagram}~(a)].
    \item If $\Omega>\frac{1}{2}$, the critical line $h_{c2}^{+}$ is located at the right of $h_{c1}^{-}$, and the phase diagram consists of six parts, corresponding the vanishment of the region V and the appearances of two regions VI and VIII [see Fig.~\ref{fig:phase-diagram}~(c)].
    \item If $\frac{\alpha}{2}<\Omega<\frac{1}{2}$,  the phase diagram consists of seven parts in which the regions V, VI, and VIII appear simultaneously [see Fig.~\ref{fig:phase-diagram}~(b)].
\end{itemize}

To verify the category of the phase transition at the critical lines, it is helpful to consider the average ground-state energy varying with the external field $h$ and the anisotropic parameter $\gamma$. Since both the quasiparticle excitation spectra satisfy $\Lambda_{k1(2)}\geq0$ for each $k>0$, the ground state is the quasiparticle vacuum state. Therefore, the ground-state energy density is given by
\begin{equation}
    e_{0} = -\frac{1}{N}\sum_{k}(\Lambda_{k1}+\Lambda_{k2}).
\end{equation}
In Figs.~\ref{fig:average-energy-h} and \ref{fig:average-energy-gamma}, we display the second-order derivatives of $e_{0}$ with respect to $h$ and $\gamma$, respectively. The divergence of the second-order derivative $\frac{d^{2}e_{0}}{dh^{2}} (\frac{d^{2}e_{0}}{d\gamma^{2}})$ at the critical point of phase transition indicates that the system undergoes a second-order phase transition.

Subsequently, we will determine the characteristics of each phase by calculating the order parameters within each phase.

\section{Order parameters}

\begin{figure}
    \centering
    \includegraphics[width=1\linewidth]{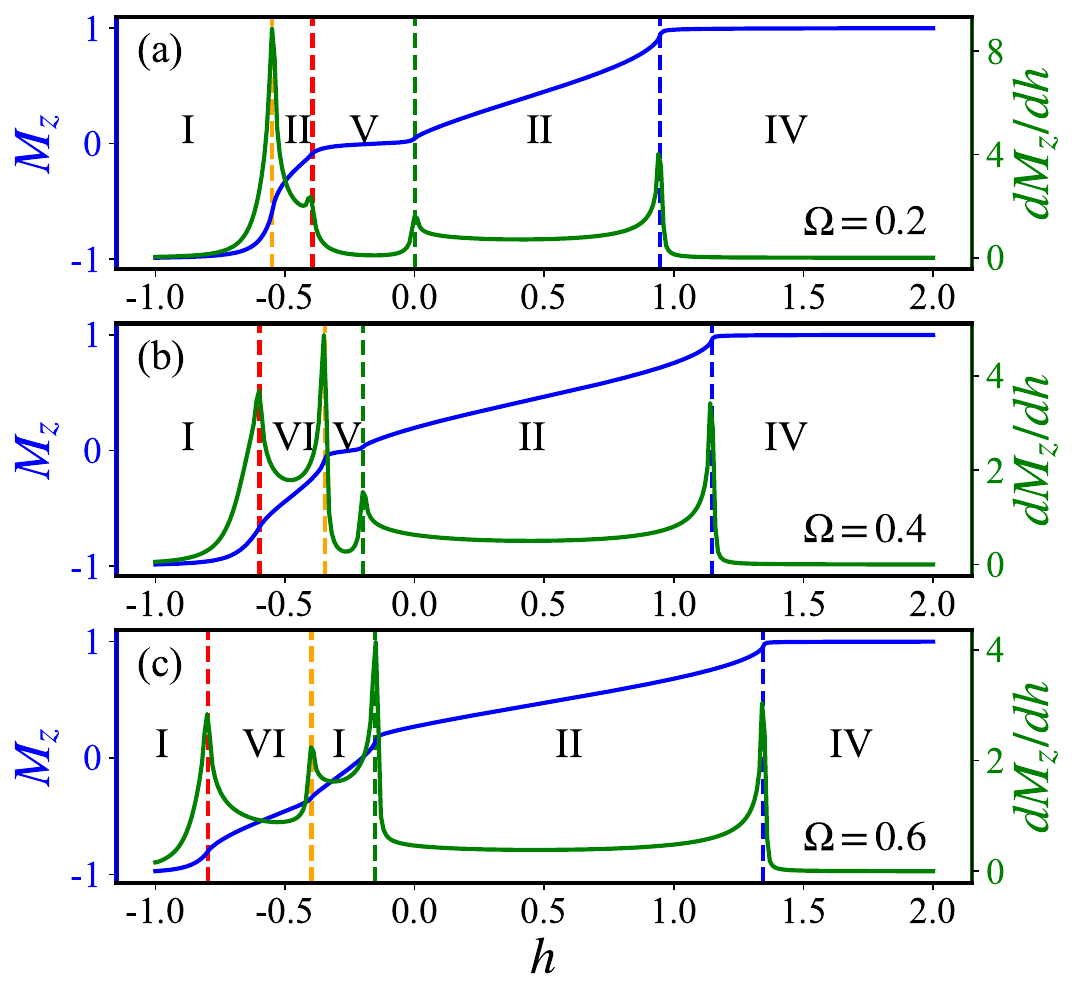}
    \caption{The average magnetization $M_{z}$ and its first-order derivative $\frac{dM_{z}}{dh}$ as functions of the external field $h$ for (a) $\Omega=0.2$, (b) $\Omega=0.4$, and (c) $\Omega=0.6$, with fixed $\alpha=0.5$ and $\gamma=0.2$.   }
    \label{fig:average-magnet}
\end{figure}

Due to the invariance of the Hamiltonian under the rotation transformation $(\sigma^{x}, \sigma^{y}, \sigma^{z}) \rightarrow (-\sigma^{x}, -\sigma^{y}, \sigma^{z})$, the order parameters $m^{x}=\langle\psi_{0}|\sigma_{n}^{x}|\psi_{0}\rangle$ and $m^{y}=\langle\psi_{0}|\sigma_{0}^{y}|\psi_{0}\rangle$ must give zero. Instead, the order parameter $m_{z}=\langle\psi_{0}|\sigma_{n}^{z}|\psi_{0}\rangle$ and the correlation functions $C_{r}^{x}=\langle\psi_{0}|\sigma_{n}^{x}\sigma_{n+r}^{x}|\psi_{0}\rangle$, and $C_{r}^{y}=\langle\psi_{0}|\sigma_{n}^{y}\sigma_{n+r}^{y}|\psi_{0}\rangle$ are better measures of the long-range order. Following the method used by Lieb et al. \cite{Lieb1961407}, we can express the spin-spin correlators in terms of fermionic operators, such as
\begin{equation}
    \begin{split}
        C_{r}^{x} & = \langle\psi_{0}|\sigma_{n}^{x}\sigma_{n+r}^{x}|\psi_{0}\rangle \\
                  & = \langle\psi_{0}|(c_{n}^{\dag}+c_{n})\exp{(i\pi\sum_{m=n}^{n+r-1}c_{m}^{\dag}c_{m})}(c_{n+r}^{\dag}+c_{n+r}) |\psi_{0}\rangle.
    \end{split}
\end{equation}
Next, using Wick's theorem, the correlation function is given by a determinant
\begin{equation}
    C_{r}^{x} = \left|\begin{array}{cccc}
                        G_{n,n+1} & G_{n,n+2} & \cdots & G_{n,n+r} \\
                        \vdots & \vdots & \vdots & \vdots \\
                        G_{n+r-1,n+1} & G_{n+r-1,n+2} & \cdots & G_{n+r-1,n+r}
                      \end{array}\right|,
\end{equation}
with
\begin{equation} \label{eq:G-elemet}
    G_{m,n} = \langle (c_{m}^{\dag}-c_{m})(c_{n}^{\dag}+c_{n}) \rangle.
\end{equation}
Similarly, we obtain
\begin{equation}
    C_{r}^{y} = \left|\begin{array}{cccc}
                        G_{n+1,n} & G_{n+1,n+1} & \cdots & G_{n+1,n+r-1} \\
                        \vdots & \vdots & \vdots & \vdots \\
                        G_{n+r,n} & G_{n+r,n+1} & \cdots & G_{n+r,n+r-1}
                      \end{array}\right|,
\end{equation}
and
\begin{equation}
    m^{z} = 2\langle c_{n}^{\dag}c_{n} \rangle-1 = G_{nn}.
\end{equation}

\begin{figure}
    \centering
    \includegraphics[width=\linewidth]{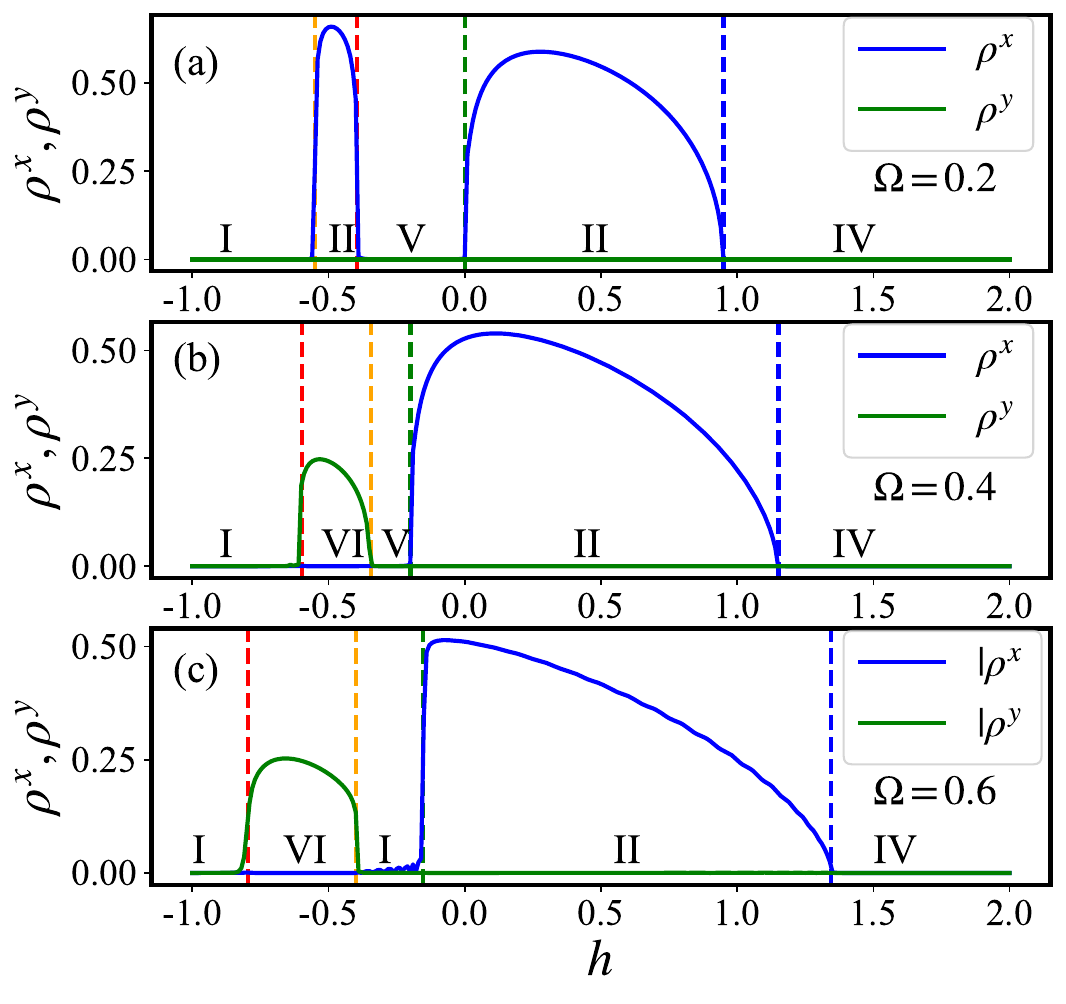}
    \caption{The long-range order parameters $\rho^{x}=C_{\frac{N}{2}}^{x}$ and $\rho^{y}=C_{\frac{N}{2}}^{y}$ as functions of the external field $h$ for (a) $\Omega=0.2$, (b) $\Omega=0.4$, and (c) $\Omega=0.6$, with fixed $\alpha=0.5$ and $\gamma=0.2$. }
    \label{fig:coxy-h}
\end{figure}

In Fig.~\ref{fig:average-magnet}, we display the average magnetization
\begin{equation}
    M_{z} = \frac{1}{N}\sum_{n}m_{n}^{z} = \frac{1}{N}\sum_{n}G_{nn}
\end{equation}
as a function of the external field $h$ for the three typical cases as shown in Fig.~\ref{fig:average-energy-h} and \ref{fig:average-energy-gamma}. From Figs.~(a) and (b), it can be observed that the average magnetization $M_{z}$ exhibits plateaus (remain constant with the increase of $h$) at phases I, V, and IV, corresponding to $M_{z} = -1$ in phase I ($h<h_{c2}^{-}$), $M_{z} = 1$ in phase IV, and a finite value approaching zero ($M_{z} \approx 0$) in phase V. It is evident that phases I and IV exhibit the paramagnetic behavior here. Additionally, the first-order derivatives $\frac{dM_{z}}{dh}$ of the average magnetization show the nonanalytic singularities at the critical points between each phase.

\begin{figure}
    \centering
    \includegraphics[width=0.95\linewidth]{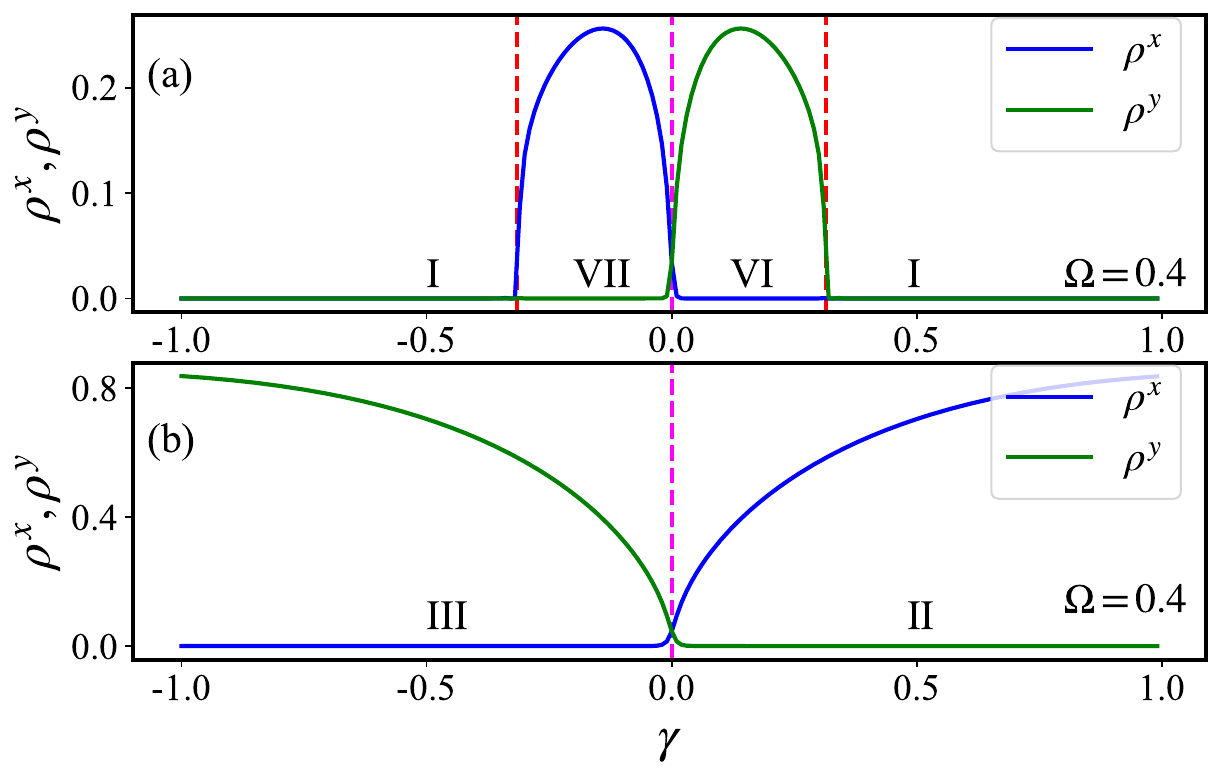}
    \caption{The long-range order parameters $\rho^{x}=C_{\frac{N}{2}}^{x}$ and $\rho^{y}=C_{\frac{N}{2}}^{y}$ as functions of the anisotropic parameter $\gamma$ for (a) $h=-0.5$, and (b) $h=0.5$, with fixed $\alpha=0.5$ and $\Omega=0.4$.}
    \label{fig:coxy-gamma}
\end{figure}

Since the order parameters $m^{x}$ and $m^{y}$ equal zero all the time, we consider the long-range order characterized by
\begin{eqnarray}
    \rho^{x} &=& \lim_{r\rightarrow\infty}C_{r}^{x}, \\
    \rho^{y} &=& \lim_{r\rightarrow\infty}C_{r}^{y}.
\end{eqnarray}
For the system with the periodic boundary condition, the limitation of $r$ takes $\frac{N}{2}$, corresponding to $C_{\frac{N}{2}}^{x}$ and $C_{\frac{N}{2}}^{y}$. In general, the phase with $\rho^{x}=0$ and $\rho^{y}=0$ is referred to as a PM phase while the phase with $\rho^{x}\neq0$ (or $\rho^{y}\neq0$) is referred to as a ferromagnetic phase along $x$ direction (or $y$ direction).
In Fig.~\ref{fig:coxy-h}, we display the long-range order parameter $\rho^{x}=C_{\frac{N}{2}}^{x}$ and $\rho^{y}=C_{\frac{N}{2}}^{y}$ as functions of the external field $h$. As seen in Fig.~\ref{fig:coxy-h}~(a), $\rho^{x}$ has a finite value in phase II but vanishes in other phases like $\rho^{y}$, which indicates that the phase II exhibits the ferromagnetic behavior along the $x$ direction. However interestingly, for the cases with additional phases VI and VII shown in Fig.~\ref{fig:average-energy-h}~(b) and (c), $\rho^{y}$ is found to have a finite value in phase VII and $\rho^{x}$ equals zero simultaneously. This suggests that the additional phase VI may take the ferromagnetic behavior along the $y-$direction. To further discuss the properties of phases VI and VII, we show the results that $\rho^{x}=C_{\frac{N}{2}}^{x}$ and $\rho^{y}=C_{\frac{N}{2}}^{y}$ vary with the anisotropic parameter $\gamma$ with fixed $h=-0.5$ (crossing the critical line between phase VI and VIII) in Fig.~\ref{fig:average-energy-gamma}~(a). We also compare the results for $h=0.5$, which crosses the critical line between phases II and III, in Fig.~\ref{fig:coxy-gamma}~(b). It is clear that from Fig.~\ref{fig:coxy-gamma}~(b) phases II and III show the characteristics of the ferromagnetic phase \cite{Zhong201043}. Similarly, we confirm that phase VI takes the ferromagnetic behavior along the $y-$ direction and phase VII takes the ferromagnetic behavior along the $x-$ direction.

\begin{table}
  \caption{A outline of the order parameters $M_{z}$, $\rho^{x}$, and $\rho^{y}$ in each phase.}
  \label{tab:order-parameter}
  \begin{ruledtabular}
    \begin{tabular}{cccc}
            & $M_{z}$ & $\rho^{x}$ & $\rho^{y}$ \\
        \hline
        I   & -1 ($h<h_{c2}^{-}$) & 0 & 0 \\
        \hline
        II  & change with $h$ & $>0$ & 0 \\
        \hline
        III & change with $h$ & 0 & $>0$ \\
        \hline
        IV  & 1 & 0 & 0 \\
        \hline
        V   & $\approx0$ & 0 & 0 \\
        \hline
        VI  & change with $h$ & 0 & $>0$ \\
        \hline
        VII & change with $h$ & $>0$ & 0 \\
    \end{tabular}
  \end{ruledtabular}
\end{table}

In a brief summary, we outline the behavior of the order parameters $M_{z}$, $\rho^{x}$, and $\rho^{y}$ in Table.~\ref{tab:order-parameter}. We notice that except for phase V, other phases are easy to distinguish: phases I and IV are the PM phase, phases II and VII are the FM phase along the $x$ direction, and phases III and VI are the FM phase along the $y$ direction (FM$_y$).

As for phase V, the results show it has anti-magnetic behavior. We further study the nearest-neighbor transverse correlation functions within and between supercells, i.e. $C_{2l-1,2l}^{z}=\langle\psi_{0}|\sigma_{2l-1}^{z}\sigma_{2l}^{z}|\psi_{0}\rangle$ and $C_{2l,2l+1}^{z}=\langle\psi_{0}|\sigma_{2l}^{z}\sigma_{2l+1}^{z}|\psi_{0}\rangle$. We consider the average nearest-neighbor transverse correlation (TC) functions
\begin{eqnarray}
    TC_{\mathrm{odd}} &=& \frac{2}{N}\sum_{l=1}^{N/2}C_{2l-1,2l}^{z}, \\
    TC_{\mathrm{even}} &=& \frac{2}{N}\sum_{l=1}^{N/2}C_{2l,2l+1}^{z},
\end{eqnarray}
where the local transverse spin-spin correlation function is obtained by
\begin{equation}
    C_{n,n+r}^{z} = G_{nn}G_{n+r,n+r} - G_{n,n+r}G_{n+r,n}.
\end{equation}
Note that $TC_{\mathrm{odd}}$ denotes the correlation of spins connected with the strong hopping interaction $J_{0}+J_{1}$ within a supercell, and $TC_{\mathrm{even}}$ denotes the correlation of spins connected with weak hopping interaction $J_{0}-J_{1}$ between the near-neighbor supercells.

\begin{figure}
    \centering
    \includegraphics[width=\linewidth]{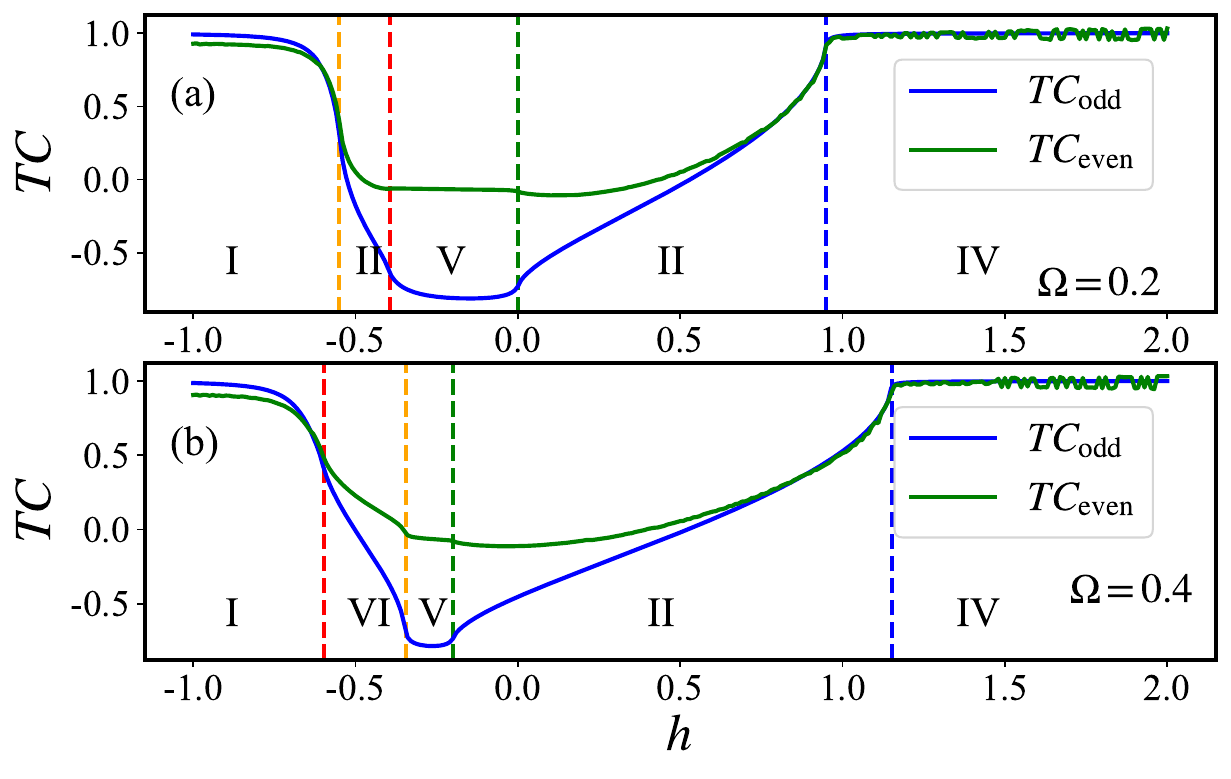}
    \caption{The average transverse spin-spin correlation functions $TC_{odd}$ and $TC_{even}$ as functions of the external field $h$ for (a) $\Omega = 0.2$, and for (b) $\Omega = 0.4$, with fixed $\alpha=0.5$ and $\gamma=0.2$. }
    \label{fig:coz}
\end{figure}

In Fig.~\ref{fig:coz}, we display the average transverse spin-spin correlation functions $TC_{\mathrm{odd}}$ and $TC_{\mathrm{even}}$ as functions of the external field $h$. It is observed that both $TC_{\mathrm{odd}}$ and $TC_{\mathrm{even}}$ converge towards the unit for systems in phases I and IV, indicating all spins polarized along the direction of external fields. However, for systems in phase V, we see $TC_{\mathrm{odd}}$ converge towards a negative value approaching the minus unit, i.e. $\min(TC_{\mathrm{odd}})\approx-0.89$, but $TC_{\mathrm{even}}$ approaches approximately zeros. This indicates that the spins within a supercell act equivalent to a spinon with a small total spin, but between the supercells are distributed randomly. Similarly, as a supplement, we show the results of the more dimerized case with $\alpha=0.1$ in Appendix~C. We observe $TC_{\mathrm{odd}}\approx-0.998$ in the more dimerized case, which implies that the spins within a supercell tend to point to opposite directions of the external field in the dimerized limit case.
This is exactly the reason for the average magnetization $M_{z}$ vanishing in phase V.
As a result, the system within phase I ends in the paramagnetic states like $|\leftarrow\leftarrow...\leftarrow\rangle$, within phase IV like $|\rightarrow\rightarrow...\rightarrow\rangle$, and within phase V like $|...\leftarrow\rightarrow\rightarrow\leftarrow...\rangle$. Therefore, phase V is essentially different from the paramagnetic phases.

\section{Quantum entanglement}

\begin{figure}
    \centering
    \includegraphics[width=1\linewidth]{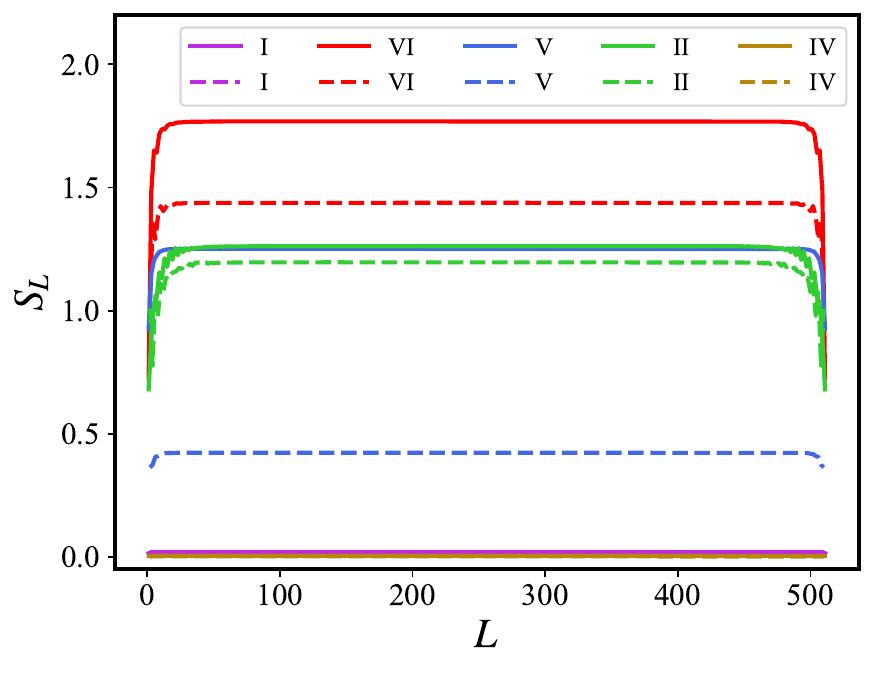}
    \caption{The von Neumann entropy as a function of the size $L$ of the sublattice for the system within each phase. The system parameters for each phase are $h=-1.0$ for phase I, $h=-0.5$ for phase VI, $h=-0.3$ for phase V, $h=0.5$ for phase II, and $h=1.5$ for phase IV, with fixed $\alpha=0.5$, $\Omega=0.4$, and $\gamma=0.2$. The solid lines show the results of the odd $L$, and the short-dashed lines show that of the even $L$. The system size here is $N=512$.}
    \label{fig:entropy-in-phase}
\end{figure}

With the development in the study of quantum computation and quantum information, the quantum entanglement has drawn increasing interest \cite{Nielsen2011, Amico200880}. In the context of statistical physics, the connection between entanglement and QPT is widely studied, where the entanglement increases as the system size scaling as $S_{L}\sim A\log{L}$ \cite{Vidal200390, Calabrese200406, Its200538, Peschel200412}. The coefficient $A$ is found to depend on the type of phase transition \cite{Zhong201043}. Specifically, for the Ising transition, $S_{L}$ is proportional to $\frac{1}{6}\log_{2}{L}$, and for the anisotropic transition, $S_{L}$ is proportional to $\frac{1}{3}\log_{2}{L}$. In the following, we will discuss the entanglement in our model to gain a deeper understanding of the relationship between entanglement and quantum phase transitions. By studying the scaling behavior of entanglement in our model, we hope to provide further confirmation of the different types of phases.

The entanglement is basically measured by the von Neumann entropy, which is defined as the negative average of the logarithm of the eigenvalues of the reduced density matrix of a subsystem of the system and is given by
\begin{equation}
    S_{L} = -\mathrm{tr}(\rho_{L}\log_{2}{\rho_{L}}).
\end{equation}
Here, $\rho_{L}$ is the reduced density matrix for $L$ contiguous spins. The von Neumann entropy $S_{L}$ can be written as \cite{Vidal200390}
\begin{equation}
    \begin{split}
        S_{L} & = -\sum_{n=1}^{L}[\frac{1-\lambda_{n}}{2}\log_{2}{(\frac{1-\lambda_{n}}{2})} \\
              & \quad\quad\quad + \frac{1+\lambda_{n}}{2}\log_{2}{(\frac{1+\lambda_{n}}{2})}],
    \end{split}
\end{equation}
where $\lambda_{n}$ are all positive eigenvalues of the matrix
\begin{equation}
    \Lambda = \left(
                \begin{array}{cccc}
                    \Pi_{11} & \Pi_{12} & \cdots & \Pi_{1L} \\
                    \Pi_{21} & \Pi_{22} & \cdots & \Pi_{2L} \\
                    \cdots   & \cdots   & \cdots & \cdots \\
                    \Pi_{L1} & \Pi_{L2} & \cdots & \Pi_{LL} \\
                \end{array}
              \right)
\end{equation}
with
\begin{equation}
    \Pi_{mn} = \left(
                \begin{array}{cc}
                    0       & G_{mn} \\
                    -G_{nm} & 0 \\
                \end{array}
               \right),
\end{equation}
where $G_{mn}$ is defined in Eq.~(\ref{eq:G-elemet}).

\begin{figure}
    \centering
    \includegraphics[width=1\linewidth]{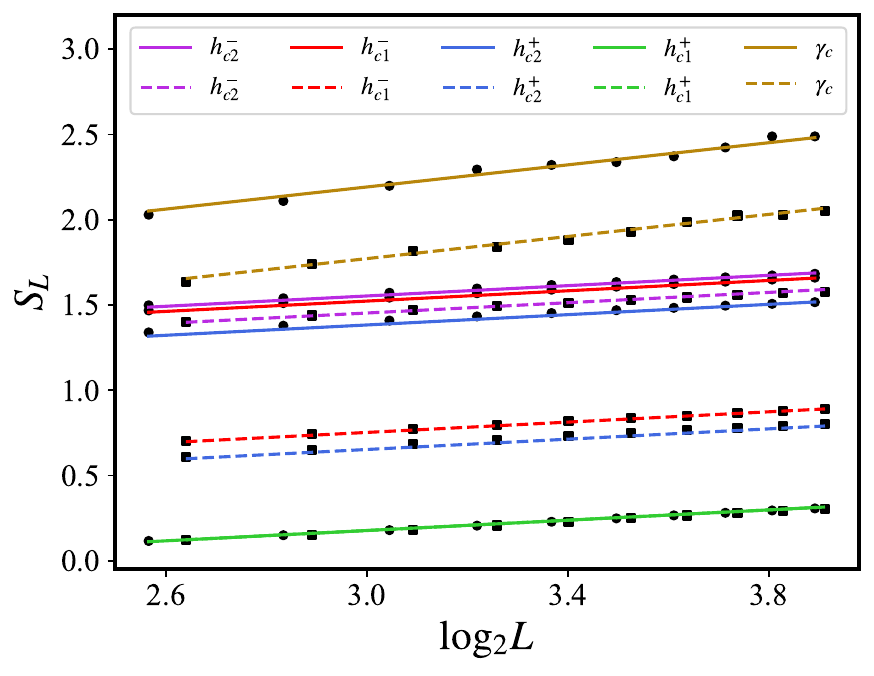}
    \caption{The von Neumann entropy as a function of $\log_{2}{L}$ for the system at the critical point. Four typical results are calculated for the critical points are $h_{c2}^{-}=-0.6$, $h_{c1}^{-}\approx0.348$, $h_{c2}^{+}=-0.2$, and $h_{c1}^{+}\approx1.148$ with fixed $\alpha=0.5$, $\Omega=0.4$, and $\gamma=0.2$. The slopes of these cases from the numerical results are 
    approaching $\frac{1}{6}$. The other one is calculated for the critical point $\gamma_{c}=0$ with $\alpha=0.5$, $\Omega=0.4$, and $h=-0.5$. The slope is 
    approaching $\frac{1}{3}$. The solid lines with the circle markers show the results of the odd $L$, and the dashed lines with square markers show that of the even $L$.  }
    \label{fig:entropy-critical-point}
\end{figure}

In Fig.~\ref{fig:entropy-in-phase}, we display the von Neumann entropy as a function of the size $L$ for the system within each phase. For simplicity, we only show the results of the system with $\Omega=0.4$, for which phases V, VI, and VII appear simultaneously. Since the definition of von Neumann entropy is based on the local block in the finite chain, the values of von Neumann entropy are observed to be different for the odd and even $L$. For each case shown in Fig.~\ref{fig:entropy-in-phase}, we can see that the von Neumann entropy has a small value approaching zero in phases I and IV. This indicates that the states in phases I and IV are deterministic, consistent with the behavior of the PM phase where all spins point to the direction of the external field. However, the von Neumann entropy in phase V has certainly large value, suggesting its difference from that in the PM phase. The result is easy to explain that the spins between the supercells are distributed randomly in phase V so that the system possesses large von Neumann entropy.

Then, we consider the von Neumann entropy near the quantum critical point. Fig.~\ref{fig:entropy-critical-point} exhibits the von Neumann entropy as a function of $\log_{2}{L}$ for the system at the critical point. Although the von Neumann entropy has different values for the odd $L$ and even $L$, the scaling behavior of it varying with the size $L$ is the same. The numerical results show that the von Neumann entropy scales as $S_{L}\sim\frac{1}{6}\log_{2}{L}$ for the system at the critical points $h_{c1}^{\pm}$ and $h_{c2}^{\pm}$. This suggests that the system undergoes the Ising transition by changing the external field $h$ across the critical points. While for the system at the critical points $\gamma_{c}=0$, we observe that the von Neumann entropy scales as $S_{L}\sim\frac{1}{3}\log_{2}{L}$. Similarly, this suggests that the system undergoes the anisotropic transition by changing the anisotropic parameter $\gamma$.

\section{Conclusion}

\begin{figure}
    \centering
    \includegraphics[width=1\linewidth]{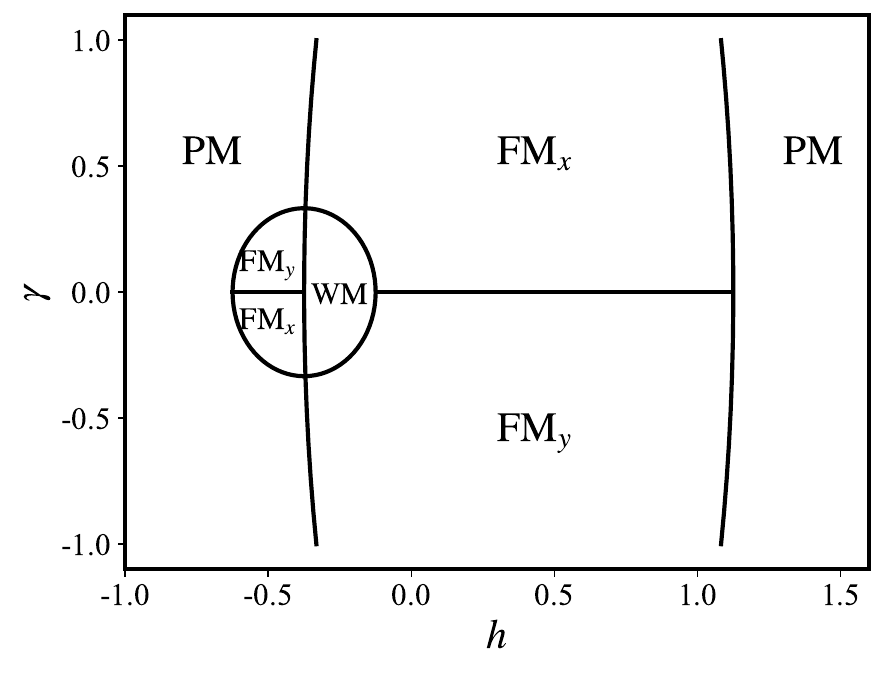}
    \caption{The final phase diagram in the ($h-\gamma$) plane for $\alpha=0.5$ and $\Omega=0.37$, which is the intermediate case with all phases coexisting.}
    \label{fig:phase-diagram-final}
\end{figure}

In this paper, we investigate the quantum phase transition in the alternating XY chain with XZX+YZY type of three-spin interactions. By obtaining the exact analytic expressions of the quasiparticle excitation spectra, we find that the quantum critical lines satisfy the inverse symmetry with respect to the points $(\pm\Omega, 0)$ in the $(h, \gamma)$ plane, where $\Omega$ is the strength of the three-site interaction. Three different types of ground state phase diagrams are thus immediately discovered. By examining the behavior of the average ground state energy, we verify that the system undergoes second-order QPTs governed by the external field and the anisotropic parameter. The intermediate case with all phases coexisting is seen in Fig.~\ref{fig:phase-diagram-final}, where the phase diagram consists of seven regions: two PM phases, two FM$_x$ phases, two FM$_y$ phases, and a WM phase. Each phase is identified by calculating the average magnetization, long-range two-point order parameters, and von Neumann entropy. Except for the WM phase, other phases have been studied very thoroughly. By calculating the nearest-neighbor transverse correlator $C_{r}^{z}$, we find that the spins within a supercell construct a cluster spinon with a small total spin, but between the supercells are not coupled. Especially for the dimerized limit case, it can be inferred that the spins tend to point to the opposite directions of the transverse field, in which the magnetization will vanish. Our results explain the reason why weak magnetic phases (and even the zero-magnetic phase) appear under alternating modulation, which has not been well understood before.

\begin{acknowledgments}
  The work is supported by the National Science Foundation of China (Grant Nos.~12204432, 11975126), and Key Research Projects of Zhejiang Lab (Nos. 2021PB0AC01 and 2021PB0AC02).
\end{acknowledgments}

\appendix

\section{Diagonalization of the Hamiltonian}

By implementing the Jordan-Wigner transformation, we transform the Hamiltonian (\ref{eq: Hamil}) into the spinless free fermion model in the double lattice by
\begin{equation}
    \begin{split}
        H & = -\frac{1}{2}\sum_{n=1}^{N'}[c_{n,1}^{\dag}c_{n,2}+ \gamma c_{n,1}^{\dag}c_{n,2}^{\dag} + h.c.] \\
          &   \quad -\frac{1}{2}\sum_{n=1}^{N'}[\alpha(c_{n,2}^{\dag}c_{n+1,1}+ \gamma c_{n,2}^{\dag}c_{n+1,1}^{\dag} + h.c.)] \\
          &   \quad -\frac{1}{2}\sum_{n=1}^{N'}[h(c_{n,1}^{\dag}c_{n,1}+ c_{n,2}^{\dag}c_{n,2} + h.c.)] \\
          &   \quad +\frac{1}{2}\sum_{n=1}^{N'}[\Omega(c_{n,1}^{\dag}c_{n+1,1} + c_{n,2}^{\dag}c_{n+1,2} + h.c.)]
    \end{split}
\end{equation}
with $N'=N/2$. The periodic boundary condition demands that $c_{N+1,1(2)}^{\dag}=c_{1,1(2)}^{\dag}$. After performing the Fourier transformation with a unit lattice constant, i.e. $c_{n,1(2)}^{\dag} = \frac{1}{\sqrt{N'}}\sum_{k}c_{k1(2)}^{\dag}e^{ikn}$ and $c_{n,1(2)} = \frac{1}{\sqrt{N'}}\sum_{k}c_{k1(2)}e^{-ikn}$, 
we can express the above Hamiltonian in the form
\begin{equation}
    H = \sum_{k>0} H_{k} = \sum_{k>0} \Psi_{k}^{\dag}\mathbb{H}_{k}\Psi_{k}
\end{equation}
in the half Brillouin zone $k\in(0,\pi]$, where the spinor operator is $\Psi_{k}^{\dag} = (c_{k1}^{\dag}, c_{k2}^{\dag}, c_{-k1}, c_{-k2})$, and the Bloch Hamiltonian is in the matrix form
\begin{widetext}
    \begin{equation}
        \mathbb{H}_{k} = \left(
                            \begin{array}{cccc}
                                -(h-\Omega\cos{k}) & 0 & -\frac{1}{2}(1+\alpha e^{ik}) & -\frac{1}{2}\gamma(1-\alpha e^{ik}) \\
                                0 & h-\Omega\cos{k} & \frac{1}{2}\gamma(1-\alpha e^{ik}) & \frac{1}{2}(1+\alpha e^{ik}) \\
                                -\frac{1}{2}(1+\alpha e^{-ik}) & \frac{1}{2}\gamma(1-\alpha e^{-ik}) & -(h-\Omega\cos{k}) & 0 \\
                                -\frac{1}{2}\gamma(1-\alpha e^{-ik}) & \frac{1}{2}(1+\alpha e^{-ik}) & 0 & h-\Omega\cos{k} \\
                            \end{array}
                         \right).
    \end{equation}
\end{widetext}
The Bloch Hamiltonian $\mathbb{H}_{k}$ can be diagonalized by the standard diagonalization procedure of the Hermitian matrix, which yields
\begin{equation}
    \mathbb{H}_{k} = U\Lambda_{k}U^{\dag}
\end{equation}
with $\Lambda_{k} = \mathrm{diag}(\Lambda_{k1}, \Lambda_{k2}, -\Lambda_{-k1}, -\Lambda_{-k2})$ and $UU^{\dag}=I$. By defining the following canonical transformation
\begin{equation}\label{eq:canonical-transformation}
    (\eta_{k1}^{\dag}, \eta_{k2}^{\dag}, \eta_{-k1}, \eta_{-k2}) = (c_{k1}^{\dag}, c_{k2}^{\dag}, c_{-k1}, c_{-k2})U,
\end{equation}
we finally obtain the Hamiltonian in the diagonal form
\begin{equation}
    H = \sum_{k}[\Lambda_{k1}(\eta_{k1}^{\dag}\eta_{k1}-\frac{1}{2}) + \Lambda_{k2}(\eta_{k2}^{\dag}\eta_{k2}-\frac{1}{2})].
\end{equation}

\section{Eigenvalues of the Bloch Hamiltonian}

The eigenvalue of the Bloch Hamiltonian $\mathbb{H}_{k}$ can be obtained by solving a corresponding quartic equation. From the eigenequation
\begin{equation}
    \mathbb{H}_{k}\Phi_{k} = \varepsilon_{k}\Phi_{k},
\end{equation}
we obtain the following quartic equation concerning the eigenvalue
\begin{align}
    \varepsilon_{k}^{4} + \varepsilon_{k}^{2}[-2|A_{k}|^{2}-2|B_{k}|^{2}-2(h-\Omega\cos{k})^{2}] \\
    (h-\Omega\cos{k})^{4}+2(h-\Omega\cos{k})^{2}(|B_{k}|^{2}-|A_{k}|^{2}) \\
    + (B_{k}^{2}-A_{k}^{2})[(B_{k}^{*})^{2}-A_{k}^{*})^{2}] = 0
\end{align}
with
\begin{equation}
    A_{k} = \frac{1}{2}(1+\alpha e^{ik}), \quad \text{and} \quad B_{k} = \frac{1}{2}\gamma(1-\alpha e^{ik}).
\end{equation}
By setting
\begin{equation}
    P = 2|A_{k}|^{2} + 2|B_{k}|^{2} + 2(h-\Omega\cos{k})^{2},
\end{equation}
\begin{equation}
    \begin{split}
        Q & = (h-\Omega\cos{k})^{4} + 2(h-\Omega\cos{k})^{2}(|B_{k}|^{2}-|A_{k}|^{2}) \\
          & \quad + (B_{k}^{2}-A_{k}^{2})[(B_{k}^{*})^{2}-A_{k}^{*})^{2}],
    \end{split}
\end{equation}
we simplify the quartic equation in the form
\begin{equation}
    \varepsilon_{k}^{4} + P\varepsilon_{k}^{2} + Q = 0,
\end{equation}
and obtain the solutions by
\begin{equation}
    \varepsilon_{k} = \pm\sqrt{\frac{P\pm\sqrt{P^{2}-4Q}}{2}}.
\end{equation}

\section{Method to calculate the order parameters}

The Hamiltonian  (\ref{eq: Hamil}) is turned into a quadratic form in real space
\begin{equation}
    H = \sum_{mn=1}^{N}[c_{m}^{\dag}A_{mn}c_{n}+\frac{1}{2}(c_{m}^{\dag}B_{mn}c_{n}^{\dag}+h.c.)],
\end{equation}
with
\begin{equation}
    \begin{split}
        A_{mn} & = -h\delta_{mn}-\frac{J_{n}}{2}\delta_{m,n+1}-\frac{J_{m}}{2}\delta_{m+1,n} \\
               & \quad +\frac{\Omega}{2}(\delta_{m,n+2}+\delta_{m+2,n}),
    \end{split}
\end{equation}
and
\begin{equation}
    B_{mn} = -\frac{J_{m}}{2}\gamma\delta_{m+1,n} + \frac{J_{n}}{2}\gamma\delta_{m,n+1}.
\end{equation}
Using the Bogoliubov transformation
\begin{eqnarray}
    \eta_{m} &=& \sum_{n}(\frac{\phi_{mn}+\psi_{mn}}{2}c_{n} + \frac{\phi_{mn}-\psi_{mn}}{2}c_{n}^{\dag}), \\
    \eta_{m}^{\dag} &=& \sum_{n}(\frac{\phi_{mn}+\psi_{mn}}{2}c_{n}^{\dag} + \frac{\phi_{mn}-\psi_{mn}}{2}c_{n}),
\end{eqnarray}
the Hamiltonian is reduced to the diagonal form in real space
\begin{equation}
    H = \sum_{m}\Lambda_{m}(\eta_{m}^{\dag}\eta_{m}-\frac{1}{2}),
\end{equation}
where $\Lambda_{m}$ is the excitation energy of the $m-$th quasiparticle mode. Both $\phi_{mn}$ and $\psi_{mn}$ can be obtained by solving the coupled equations
\begin{eqnarray}
    \Lambda_{m}\phi_{mn} &=& \sum_{n}\psi_{mn}(A_{nm}+B_{nm}), \\
    \Lambda_{m}\psi_{mn} &=& \sum_{n}\psi_{mn}(A_{nm}-B_{nm}).
\end{eqnarray}
By defining
\begin{eqnarray}
    U &=& \frac{\phi+\psi}{2}, \\
    V &=& \frac{\phi-\psi}{2},
\end{eqnarray}
we have
\begin{equation}
    \begin{split}
        c_{m}^{\dag}c_{n}^{\dag} & = \sum_{jj'}(U_{jm}U_{j'n}\eta_{j}^{\dag}\eta_{j'}^{\dag} + U_{jm}V_{j'n}\eta_{j}^{\dag}\eta_{j'} \\
        & \quad\quad + V_{jm}U_{j'n}\eta_{j}\eta_{j'}^{\dag} + V_{jm}V_{j'n}c_{j}c_{j'}),
    \end{split}
\end{equation}
\begin{equation}
    \begin{split}
        c_{m}c_{n} & = \sum_{jj'}(U_{jm}U_{j'n}\eta_{j}\eta_{j'} + U_{jm}V_{j'n}\eta_{j}\eta_{j'}^{\dag} \\
        & \quad\quad + V_{jm}U_{j'n}\eta_{j}^{\dag}\eta_{j'} + V_{jm}V_{j'n}c_{j}^{\dag}c_{j'}^{\dag}),
    \end{split}
\end{equation}
\begin{equation}
    \begin{split}
        c_{m}^{\dag}c_{n} & = \sum_{jj'}(U_{jm}U_{j'n}\eta_{j}^{\dag}\eta_{j'} + U_{jm}V_{j'n}\eta_{j}^{\dag}\eta_{j'}^{\dag} \\
        & \quad\quad + V_{jm}U_{j'n}\eta_{j}\eta_{j'} + V_{jm}V_{j'n}c_{j}c_{j'}^{\dag}),
    \end{split}
\end{equation}
and
\begin{equation}
    \begin{split}
        c_{m}c_{n}^{\dag} & = \sum_{jj'}(U_{jm}U_{j'n}\eta_{j}\eta_{j'}^{\dag} + U_{jm}V_{j'n}\eta_{j}\eta_{j'} \\
        & \quad\quad + V_{jm}U_{j'n}\eta_{j}^{\dag}\eta_{j'}^{\dag} + V_{jm}V_{j'n}c_{j}^{\dag}c_{j'}).
    \end{split}
\end{equation}
We thus obtain
\begin{equation}
    G_{mn} = \langle (c_{m}^{\dag}-c_{m})(c_{n}^{\dag}+c_{n}) \rangle = -\sum_{j}\psi_{jm}\phi_{jn}.
\end{equation}

\section{Case of the dimerized lattice}

\begin{figure}
    \centering
    \includegraphics[width=1\linewidth]{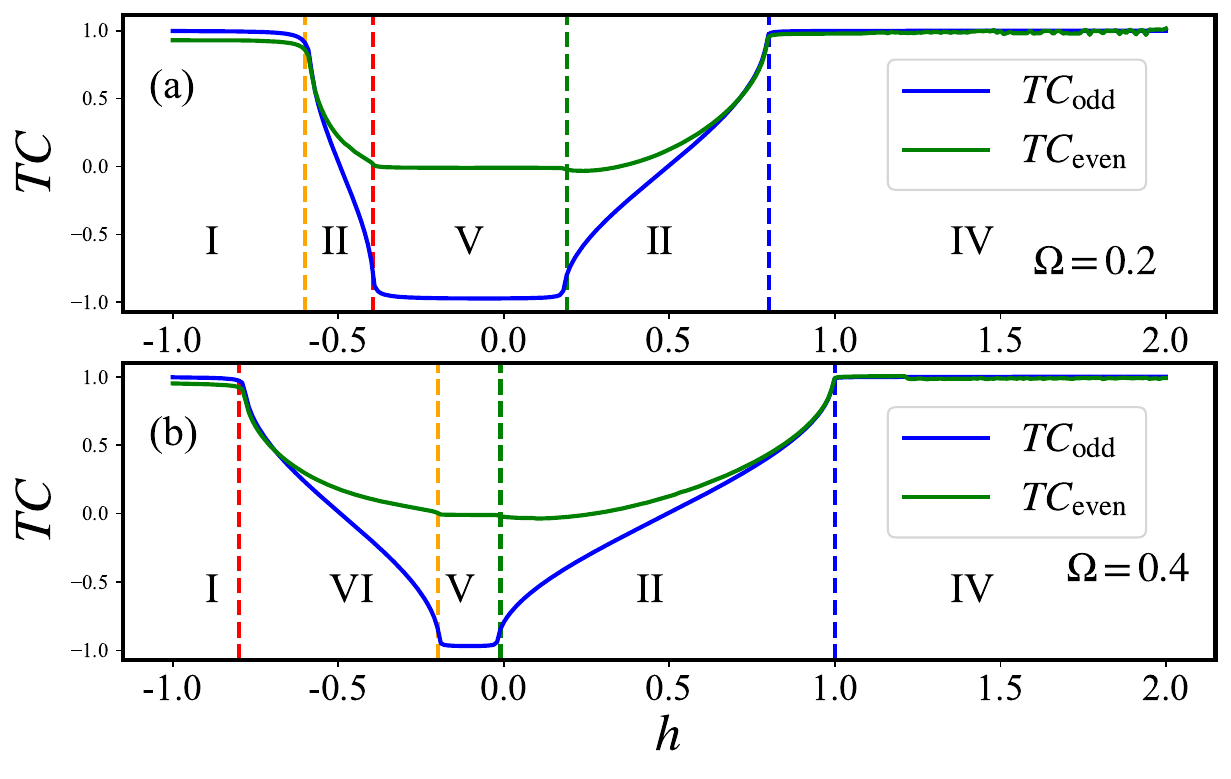}
    \caption{The average transverse spin-spin correlation functions $TC_{\mathrm{odd}}$ and $TC_{\mathrm{even}}$ as functions of the external field $h$ for (a) $\Omega = 0.2$, and for (b) $\Omega = 0.4$, with fixed $\alpha=0.2$ and $\gamma=0.2$.}
    \label{fig:coz-dimerized}
\end{figure}

Now, we show more evidence that the spins within a supercell point to the opposite direction along the external field in the dimerized limits. In Fig.~\ref{fig:coz-dimerized}, we show the average transverse spin-spin correlation functions $TC_{\mathrm{odd}}$ and $TC_{\mathrm{even}}$ as functions of the external field $h$ for $\alpha = \frac{J_{0}-J_{1}}{J_{0}+J_{1}} = 0.1$. It can be seen that, as compared to the results of $\alpha=0.5$ (see Fig.~\ref{fig:coz}), the correlation functions $TC_{\mathrm{odd}}$ and $TC_{\mathrm{even}}$ show the clearer plateaus in phase V, corresponding to $TC_{\mathrm{odd}}\approx-0.998\approx-1$ and $TC_{\mathrm{even}}\approx0$. The results indicate that in the dimerized limit, the spins within a supercell construct an anti-paramagentic cluster in phase V, and the spins between the nearest-neighbor supercells are not coupled, corresponding to the random distribution of the supercells.

\bibliography{qpt}

\end{document}